\documentclass[conference]{IEEEtran}

\usepackage[cmex10]{amsmath}
\usepackage{cite}
\usepackage{url}
\usepackage{color}

\usepackage{amssymb}
\usepackage{path}
\usepackage{verbatim}
\usepackage{algorithm}
\usepackage{framed}
\usepackage{algpseudocode}

\usepackage{color}
\usepackage{amsmath}
\usepackage{cases}
\usepackage{theorem}
\usepackage{subfig}
\usepackage{subfloat}

\usepackage{graphicx}

\usepackage{pgfplots}
\usepackage{pgfplotstable}
\pgfplotsset{compat=newest}
 
\DeclareMathOperator*{\argmaxB}{argmax} 

{ \theorembodyfont{\normalfont} 

}

\newcounter{enumctr}

\DeclareFontFamily{U}{mathx}{\hyphenchar\font45}
\DeclareFontShape{U}{mathx}{m}{n}{<-> mathx10}{}
\DeclareSymbolFont{mathx}{U}{mathx}{m}{n}
\DeclareMathAccent{\widebar}{0}{mathx}{"73}

\begin{document}

\title{An ADMM-based Optimal Transmission Frequency Management System for IoT Edge Intelligence}

\author{\IEEEauthorblockN{Hongde Wu \IEEEauthorrefmark{1}, Noel E. O'Connor \IEEEauthorrefmark{1}\IEEEauthorrefmark{2}, Jennifer Bruton \IEEEauthorrefmark{1} and Mingming Liu \IEEEauthorrefmark{1}\IEEEauthorrefmark{2}\IEEEauthorrefmark{3}}
	\IEEEauthorblockA{\IEEEauthorrefmark{1}  School of Electronic Engineering, Dublin City University, Ireland \\
	\IEEEauthorrefmark{2}  Insight Centre for Data Analytics, Dublin City University, Ireland \\
		\textit{\IEEEauthorrefmark{3} Corresponding author}: mingming.liu@dcu.ie}}
\maketitle

\begin{abstract}

In this paper, we investigate a key problem of Internet of Things (IoT) applications in practice. Our research objective is to optimize the transmission frequencies for a group of IoT edge devices under practical constraints. Our key assumption is that different IoT devices may have different priority levels when transmitting data in a resource-constrained environment and that those priority levels may only be locally defined and accessible by edge devices for privacy concerns. To address this problem, we leverage the well-known Alternating Direction Method of Multipliers (ADMM) optimization method and demonstrate its applicability for effectively managing various IoT data streams in a decentralized framework. Our experimental results show that the transmission frequency of each edge device can converge to optimality  with little delay using ADMM, and the proposed system can be adjusted dynamically when a new device connects to the system. In addition, we also introduce an anomaly detection mechanism to the system when a device's transmission frequency may be compromised by external manipulation, showing that the proposed system is robust and secure for various IoT applications. 

\end{abstract}


\IEEEpeerreviewmaketitle

\section{Introduction}

The Internet of Things (IoT) is the extension of the Internet to include connected embedded computing devices with the intention of transferring data over such a network without human-to-machine interaction \cite{madakam2015internet}. In a typical IoT set-up, data from various IoT sensors and edge devices can be produced and transmitted through gateways to the Cloud \cite{6192937}. The collected IoT data can be processed, analyzed, and stored by different cloud instances using Artificial Intelligence (AI) on cloud infrastructures. The conventional cloud-dominant centralized architecture has tremendous benefits for many IoT applications \cite{2018Brokering}, such as optimizing the cost and energy in pricing problems \cite{2018Economic, St2016Cloud} and maximizing the quality of service \cite{QUARATI2016403}. However, it also comes with inevitable shortcomings especially those involving  local users' autonomous control over their data, control that usually requires the decision making process to be conducted at the device side or edge side for better security \cite{9301390} and privacy protection \cite{9163078}.

Specifically, we now consider a typical IoT scenario where data streams from various IoT devices can be transmitted to the Cloud and stored on a cloud database.  Our initial observation is that most IoT devices start to transmit data at a fixed transmission frequency, and such a transmission frequency is typically set by default or pre-defined by the device manufacturer with limited options made available to users. However, some advanced IoT devices with edge intelligence, e.g. Raspberry Pis and the Jetson series toolkit from Nvidia, can now be programmed to promptly respond to changes in the external environment \cite{10.1117/12.2571307, 8230004}, and can also be deployed with deep learning algorithms to satisfy stringent low-latency transmission requirements for time-sensitive IoT applications \cite{9287960, 9289509}. This approach does not sufficiently cater for a practical situation where groups of IoT devices may work collaboratively with limited operational resources enforced by the external environment. In fact, implementing IoT devices in a resource-constrained environment may impose two types of design problems that are of particular interest to us in this paper: 1) how to determine an adaptive transmission frequency for each IoT device so that an overall utility of the group of devices can be maximized in response to the dynamic changes of the environment; and 2) how to ensure that different kinds of network resources can be better managed in a way that heterogeneous IoT devices can be engaged with the network in a secure, privacy-aware and plug-and-play manner.

To be specific, privacy-awareness in our context refers to the fact that the mapping between the utility and the transmission dynamics of a given IoT device should not be revealed to any unrelated devices, third-party gateway and untrusted cloud units or instances. We highlight that this design consideration is important in practice because if this information is revealed publicly it may be possible for an attacker to identify which IoT device is more vulnerable in a given system \cite{2020A}. In terms of network resources, we observe that the capacity of a cloud-based database instance is typically limited in storage space and it often comes with a capped time-based throughput for a given user. For instance, an IBM Cloudant database instance allows 1 GB of data storage with 10 writes/sec for its Lite Plan users, and 20 GB of data storage with 50 writes/sec for its Standard Plan users \cite{bienko2015ibm}. Given this scenario, it can be envisioned that a writing congestion event, e.g. a REST-API writing failure, can be triggered for a group of IoT devices if the Maximum Writing Frequency (MWF) of the data is not managed properly.

To solve this challenge, in this paper we propose a transmission frequency management system for IoT edge devices in a decentralized architecture with an anomaly detection mechanism. Thus the MWF can be managed optimally by a group of IoT devices and any abnormal writing frequency occurrences can be detected by the gateway. To carry out  optimization, we assume that each IoT device is associated with a utility function with some concavity \cite{7106504, 2019Utility}, in a way that only the user of the device can specify. Here, the utility refers to how a user can practically benefit from a given Data Flow Writing Frequency (DFWF). For instance, a utility function can easily describe the accuracy of a trained model with respect to DFWF of a given IoT device for an Edge AI type of IoT applications \cite{LV202190}. Furthermore, as previously mentioned, such a utility function may also potentially reflect the significance or vulnerability of an IoT device in a specific scenario. For instance, a faster transmission frequency of a webcam in a bank system may be more desirable, i.e., have higher utility, especially in an emergency, than that  of a $CO_2$ detector.

With this idea in mind, our main objective in this paper is to maximize the overall utility of the group of IoT devices given the predefined and limited MWF and storage capacity of the database. We will show that the presented challenge can be formulated as a concave optimization problem with constraints. This problem will then be solved using the well-known Alternating Direction Method of Multipliers (ADMM) algorithm \cite{boyd2011distributed} in a decentralized optimization framework where each utility function is locally defined on the edge device and will not be revealed to any unrelated devices and untrusted management platforms, such as other smart gateways and cloud units/instances. The proposed solution aims to provide flexibility in data transmission for IoT systems and applications especially in resource-constrained environments. As we shall see, the designed system is fully autonomous and can be easily deployed to optimally manage various IoT transmission frequencies with anomaly detection capabilities. 



The remainder of this paper is organized as follows. In Section II, the architecture of the proposed system is presented. The optimization problem is formulated in Section III and its implementation is discussed in Section IV. Establishing and configuring real-world simulations and their results are discussed in Section V. The anomaly detection mechanism is demonstrated in Section VI. Finally, a conclusion from the current research and potential future research directions are provided in Section VII.

\section{System Architecture}\label{sec:System_model}

\begin{figure}[ht]
	\vspace{-0.1in}
	\centering
	\includegraphics[width=0.5\textwidth, height=3in]{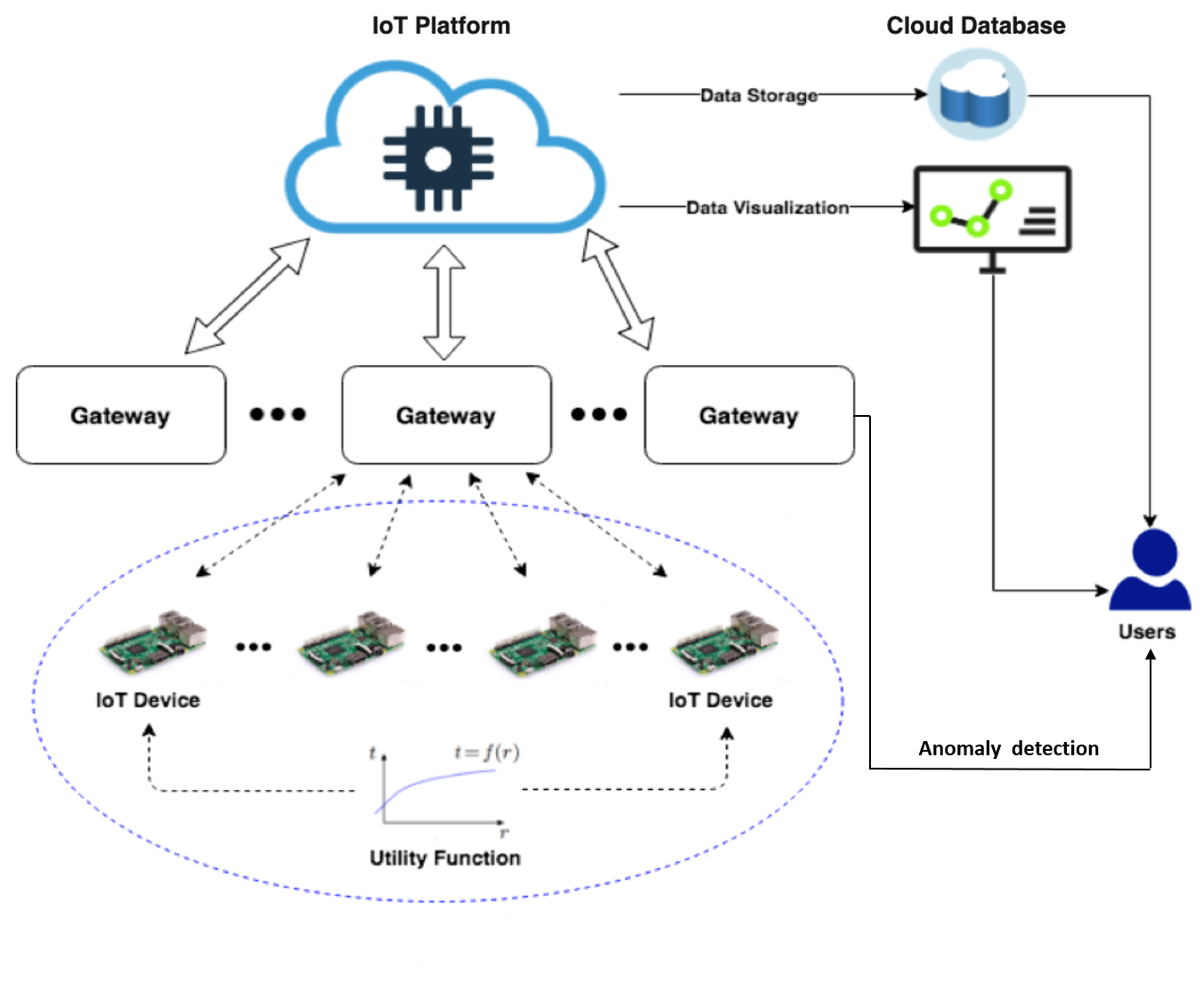}
	\caption{Schematic diagram of the system architecture.}
	\label{fig:architecture1}
	\vspace{-0.1in}
\end{figure}

Our proposed system architecture is illustrated in Fig. \ref{fig:architecture1}. The system consists of four main components, including IoT edge devices, gateways, a cloud platform and users. The main functionalities of each component are described as follows: 

\begin{itemize}
	\item[1.] \textbf{IoT devices}: sensors/devices connected to a gateway, having the capabilities of defining utility functions and the ability to solve a local optimization problem in a decentralized manner.
	\item[2.] \textbf{Gateway}: collects data from IoT devices/sensors, passing data to the Cloud, and conducts basic data processing tasks including anomaly detection to protect and inform users. 
	\item[3.] \textbf{Cloud platform}: a central hub for data analysis, monitoring and storage. 
	\item[4.] \textbf{Users}: the owner of the IoT devices who wishes to use the IoT devices in some collaborative application scenarios.
\end{itemize}

In the proposed system, a gateway starts by waiting for connection from IoT devices. When an IoT device initially connects to the gateway, the decentralized optimization algorithm is activated to calculate the optimal transmission frequencies for all connected devices whilst taking account of the resource constraints of the system. After that, the gateway starts to collect data streams from all IoT devices after the transmission frequencies are established.  Finally, data collected by the gateway is transmitted to the cloud platform for the purpose of data storage and further analysis of the IoT devices if specifically requested by the users. 

\section{Problem Statement}

We now present the specific problem statement to be solved in this paper. A user wishes to determine the optimal DFWF of every IoT edge device so that the overall utility of the whole group can be maximised, given $N$ number of devices connected to the gateway, the utility $f_i(x_i)$ of the $i^{th}$ device with current DFWF $x_i$, MWF $c$, total data storage available per received data packet, $d$, and $a_i$ the data size required for the $i$'th device per writing request.


Mathematically, this problem can be formulated as follows:

\begin{equation} \label{eq}
\begin{gathered}
\underset{x_1, x_2, \ldots, x_{N}}{\max} \quad
\sum\limits_{i = 1}^{N} f_{i}\left(x_i \right),\\
{\text{such that}} ~
\sum\limits_{i = 1}^{N} x_i \leq c,  ~ \sum\limits_{i = 1}^{N} a_i x_i \leq d, ~ x_i \geq 0  \\
\end{gathered}
\end{equation}

We shall only require that each utility function $f_i(x_i)$ can be modelled as a continuously differentiable, non-decreasing, strictly concave function, which is a common assumption for modelling the utility of internet data traffic \cite{srikant2004mathematics}. For example, utility functions may be modelled as a cluster of negative quadratic functions.

\section{System Implementation}

The classic ADMM algorithm proposed in \cite{boyd2011distributed} is particularly suited to solving the formulated optimization problem \eqref{eq} as the problem can be converted to a convex optimization problem with convex constraints. Here we briefly recall the ADMM algorithm for solving \eqref{eq}, which is shown in Algorithm \ref{ADMM}, where $x$ and $z$ are
updated in an alternating fashion and $u$ is a dual update variable.

\begin{algorithm}[htbp]
	\caption{ADMM Algorithm}
	\begin{algorithmic}[1]
		\State  $\textbf{x}^{k+1}:=  \argmaxB_\textbf{x}  ( \sum\limits_{i = 1}^{N} f_i(x_i) + (\rho/2)  || \textbf{x} - \textbf{z}^{k} + \textbf{u}^{k} ||_{2}^{2})$
		\State  $\textbf{z}^{k+1}:= \Pi_{\mathcal{C}} (\textbf{x}^{k+1} + \textbf{u}^{k})$
		\State  $\textbf{u}^{k+1}:= \textbf{u}^{k} + \textbf{x}^{k+1} - \textbf{z}^{k+1}$
	\end{algorithmic}
	\label{ADMM}
\end{algorithm}

Note that the above ADMM algorithm can be implemented in a decentralized manner as our objective function is separable which implies that both $\textbf{x}$ and $\textbf{u}$ vector updates in the algorithm can be implemented in parallel. Finally, the $\textbf{z}$ update depends on inputs from both $\textbf{x}$ and $\textbf{u}$. Given these inputs, the projection operator $\Pi_{\mathcal{C}}$ projects the resulting vector to the constrained convex space $\mathcal{C}$. Thus, the $\textbf{z}$ update needs to be implemented on a gateway. Note that $\rho$ is the augmented Lagrangian parameter and we take $\rho = 1.0$, being equivalent to a $\rho/2$ step size in $x$ update. The ADMM algorithm in its decentralized format is shown in Algorithm \ref{DecentralisedADMM}.

\begin{algorithm}[htbp]
	\caption{Decentralized ADMM Algorithm}
	\begin{algorithmic}[1]
		\State  ${x_i}^{k+1}:=  \argmaxB_{x_i}  (  f_i(x_i) + (\rho/2)  || {x_i}^{k} - {z_i}^{k} + {u_i}^{k} ||_{2}^{2})$
		\State  $\textbf{z}^{k+1}:= \Pi_{\mathcal{C}} (\textbf{x}^{k+1} + \textbf{u}^{k})$
		\State  ${u_i}^{k+1}:= {u_i}^{k} + {x_i}^{k+1} - {z_i}^{k+1}$
	\end{algorithmic}
	\label{DecentralisedADMM}
\end{algorithm}

\begin{figure}[ht]
	\vspace{-0.1in}
	\centering
	\includegraphics[width=0.4\textwidth]{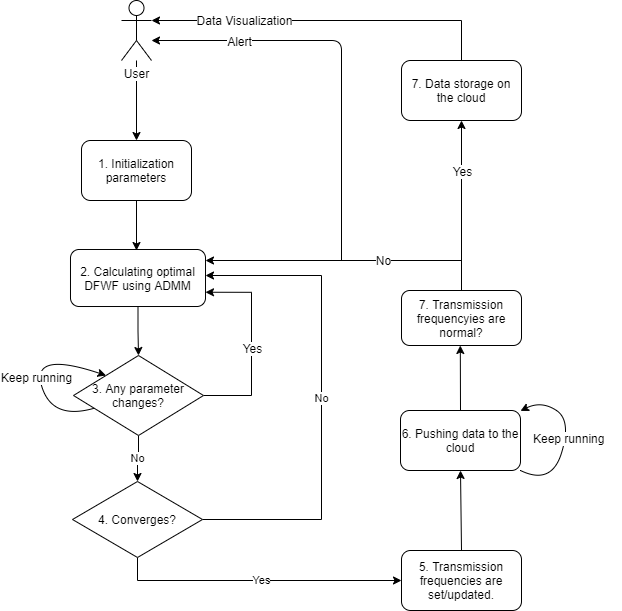}
	\caption{System implementation flowchart. }
	\label{fig:flow_chart}
	\vspace{-0.3in}
\end{figure}

With this algorithm in mind, the proposed system can be implemented in the following steps, which are illustrated in Fig. \ref{fig:flow_chart}.

\begin{itemize}
	\item[S1:] During the initialization stage, a user needs to specify some parameters before running the
	algorithm. This includes $N$, $c$, $d$, $a_i$ and the utility function $f_i(x_i)$ of each device.
	\item[S2:] When the initialization step finishes, the ADMM algorithm will be implemented in an iterative manner on the edge IoT devices to determine the optimal DFWF by computing the optimal ${x_i}^{k+1}$ as per Algorithm \ref{DecentralisedADMM}.
	\item[S3:] During each iteration, the gateway gathers all the optimal ${x_i}^{k+1}$ from all devices, calculates and  broadcasts the updated $\textbf{z}$ value to local edge devices. Upon receiving the $\textbf{z}$ value, each edge device updates ${u_i}^{k+1}$ correspondingly.
   \item [S4:] If there are any resource changes during  runtime, the algorithm can dynamically capture the changes to recalculate the optimal solution given the new context. 
	\item[S5:] When the algorithm converges, the optimal DFWF will be set by each device, and these devices
	can then start pushing data to the cloud accordingly. 
	\item[S6:] The gateway keeps monitoring the data injection and detects if an anomaly happens on any of the transmission frequencies. If so, the user will be alerted and the optimal solution will be recalculated and reset after the anomaly has been remedied.  
	\item[S7:] Finally, all transmitted data streams will be stored on the cloud and an authorised user can leverage the stored data for visualization and analysis by making a request. 
	
\end{itemize}

\begin{figure}[ht]
	\vspace{0.1in}
	\centering
	\includegraphics[width=0.4\textwidth]{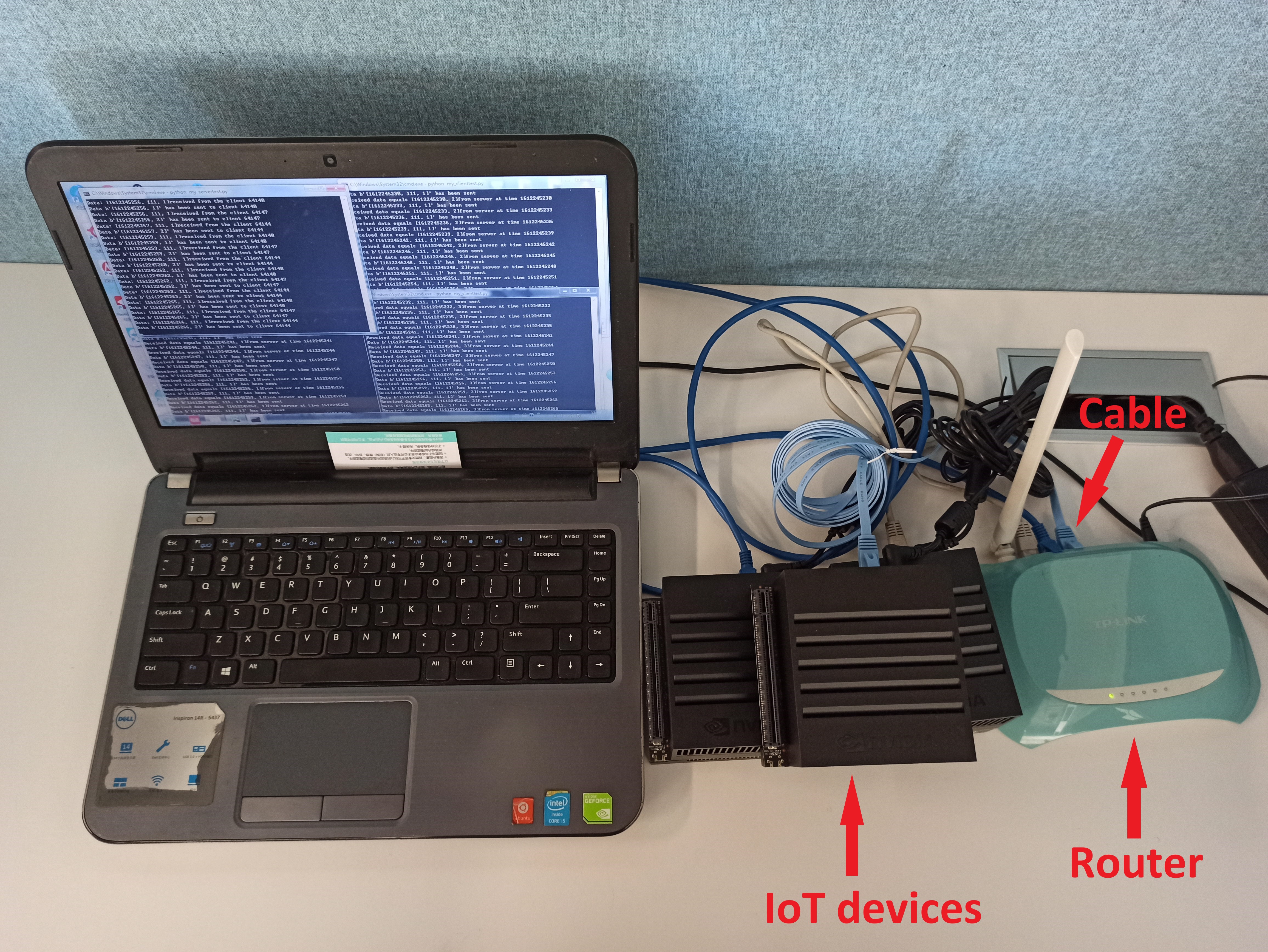}
	\caption{System simulation device setup.}
	\label{fig:devices}
	\vspace{-0.1in}
\end{figure}

\section{Simulation}
This section presents simulation results to evaluate the performance of the proposed system. As shown in Fig. \ref{fig:devices}, the system consists of a laptop as the central node (i.e., as a smart gateway in this work), three IoT devices (Jetson AGX Xavier, Nvidia), and a router for the communication between the gateway and the IoT devices. Typically, IoT s connect to router using wireless. However, in our setup, since the IoT devices do not have the capability of wireless transmission, they transmit data to the router via cables, and the laptop communicates with router wirelessly. Decentralized ADMM optimization and data transmission are implemented on both the gateway the and devices via socket programming. System parameters for the simulations are set as $N = 3$, $c = 10$, $d = 15$, $a_1 = 2$, $a_2 = 3$, and $a_3 = 5$. The utility functions in this simulation are presented in Table \ref{UtilityF} and have the characteristics previously specified to successfully apply the ADMM algorithm. We simulate the system in two scenarios: a) resources are sufficient for the data transmission request, and b) resources are insufficient for the data transmission request from all devices. For each device $i$, its transmission frequency $x_i$ is defined as data is transmitted $x_i$ times per second. In particular, $x_i = 0$ implies that the $i^{th}$ device is not transmitting data. Thus, for each device, an extra constraint, $x_i >= \gamma_i$ applies to indicate the minimum transmission frequency. For simplicity, we set $\gamma_1 = \gamma_2 = \gamma_3 = 1$ in our simulation.

It is worth noting that the gateway is not able to access the utility function of each device in order to cater for privacy concerns, and also that the transmission frequency of each device is calculated locally and not explicitly exposed to the gateway. However, a DFWF may be estimated by the gateway by evaluating the time intervals of the consecutively received data packets and an averaged DFWF is calculated over 300 data packets after the optimal DFWF is assigned.


\renewcommand\arraystretch{1.5}
\begin{table}[ht!]
	\centering
	\caption{Utility Functions}\label{UtilityF}
	\vspace{-0.2cm}
	\begin{tabular}{|c|c|}
		\hline
		\textbf{Device index} &
		\textbf{Utility Functions} \\ \hline
		1 &  $f_1(x_1) = -(x_1+9)^2 - x_1^3 + 900$ \\ \hline
		2 & $f_2(x_2) = -(x_2-4)^2 + 500$ \\ \hline
		3 & $f_3(x_3) = -(2x_3+3)^2 - x_3^3 + 110$ \\ \hline

	\end{tabular}

\end{table}

\subsection{System with sufficient resources}

\begin{figure}[ht]
	\vspace{-0.1in}
	\centering
	\includegraphics[width=0.42\textwidth]{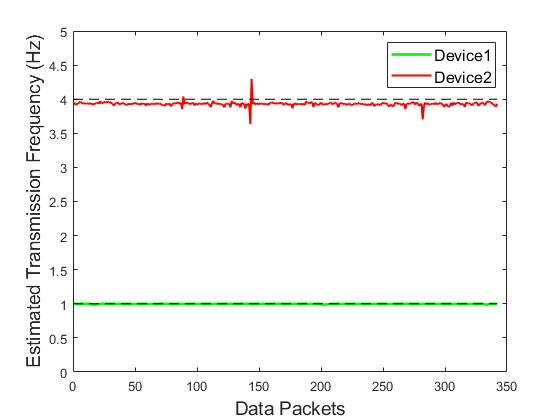}
	\caption{Decentralized optimization process of transmission frequency for Device 1 and Device 2.}
	\label{fig2a}
	\vspace{-0.1in}
\end{figure}

\begin{figure}[ht]
	\vspace{-0.1in}
	\centering
	\includegraphics[width=0.42\textwidth]{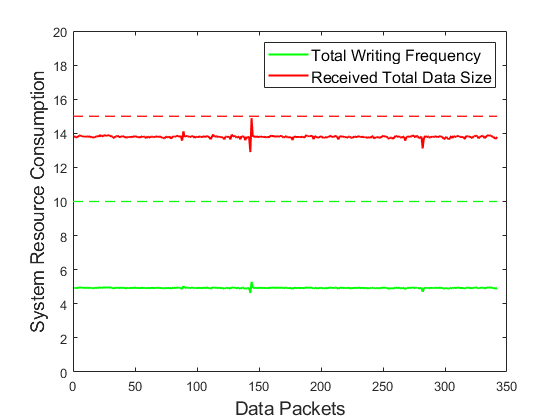}
	\caption{System resources consumption.}
	\label{fig2b}
	\vspace{-0.1in}
\end{figure}

In this scenario, only device $1$ and device $2$ are connected to the gateway (i.e., parameter $N = 2$) and all other system parameters are kept by default, i.e., $c = 10$, $d = 15$, $a_1 = 2$, $a_2 = 3$ with the associated utility functions $f_1(x_1)$ and $f_2(x_2)$ shown in Table \ref{UtilityF}. With these parameters, the theoretical optimal results of the ADMM implementation are $x_1^{*} = 1$ and $x_2^{*} = 4$ for optimization problem (1). This result implies that the gateway expects to receive $1$ and $4$ data packet(s) per second from device $1$ and $2$ on average.  In this setup, the capacity provided by the system is sufficient since $x_1^{*} + x_2^{*} < c$ and $a_1 * x_1^{*} + a_2 * x_2^{*} <d$. With the decentralized ADMM implemented using the simulation setup, the optimization results and resource consumption of the system are illustrated in Fig. \ref{fig2a} and Fig. \ref{fig2b}, respectively. In particular, Fig. \ref{fig2a} shows the evolution of the calculated DFWF for both devices as estimated by the gateway. The DFWFs are estimated along with the number of received data packets, indicated by the red and green lines for device $1$ and device $2$, respectively. Concretely, our results show that the estimated DFWFs are $0.9984\ Hz$ and $3.9318\ Hz$ for device $1$ and $2$, respectively, as shown in Table \ref{simulationA}. The estimated DFWFs are just slightly below the the theoretical optimal DFWFs, indicated by the dotted-line in Fig. \ref{fig2a}. The decay of the DFWF is accounted for by the internet speed, while the communication between the gateway and the devices is based on a router, resulting in a $1.6\ ms$ and a $4.3\ ms$ delay for device $1$ and device $2$, respectively. Meanwhile, we find that the fluctuation of the estimated DFWFs is caused by the data jamming when the gateway is receiving data packets with high writing frequency. Fig. \ref{fig2b} shows the sum of DFWFs as well as the size of total data packets of all connected devices per second transmitted to the gateway. The dotted-line indicates the maximum total DFWF (in red) and received data size (in green) for each data packet. Since the system can provide sufficient resources, the total DFWF and the writing data size has not reached the resource boundary after the transmission frequencies are optimized, indicating that the proposed system is robust as long as the system resources are sufficient for this specific data transmission task.

\begin{table}[!h]
	\centering
	\caption{Simulation results (average)}\label{simulationA}
	\vspace{-0.2cm}
	\begin{tabular}{|c|c|c|}
		\hline
		\textbf{ } & \textbf{DFWF (Hz)} & \textbf{DFWF (Hz)}  \\ \hline
		\textbf{ } & \textbf{Device 1} & \textbf{Device 2} \\ \hline
		\textbf{Theoretical} & 1.0000 & 4.0000   \\ \hline
		\textbf{Actual} & 0.9984 & 3.9318   \\ \hline
		\textbf{Absolute Error} & 0.0016 & 0.0682   \\ \hline
	\end{tabular}
\end{table}

\subsection{System with insufficient resources}

In this scenario, after device $1$ and device $2$ have connected to the gateway and the optimized transmission frequencies have been calculated, a new device, device $3$, connects to the gateway and the timing of connection is recorded. Given $N = 3$, $c = 10$, $d = 15$, $a_1 = 2$, $a_2 = 3$, $a_3 = 5$ and the corresponding utility functions $f_1(x_1)$, $f_2(x_2)$, $f_3(x_3)$ reported in Table \ref{UtilityF}, the theoretical optimal results of the ADMM implementation are calculated as  $x_1^{*} = 1.00$, $x_2^{*} = 2.66$ and $x_3^{*} = 1.00$ for optimization problem (1). This result implies that, on average, the gateway expects to receive $1$, $2.66$ and $1$ data packet(s) per second from devices $1$, $2$, and $3$ respectively.

\begin{figure}[ht]
	\vspace{-0.1in}
	\centering
	\includegraphics[width=0.41\textwidth]{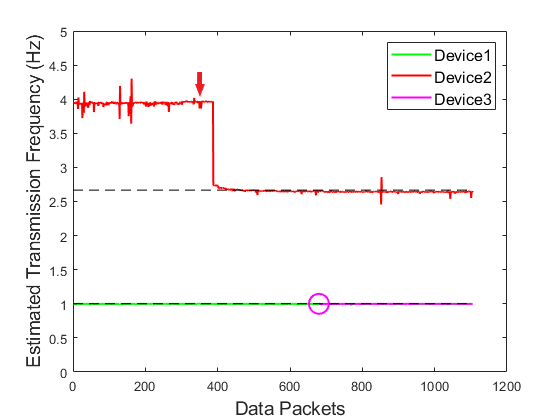}
	\caption{Decentralized optimization process of transmission frequencies for Device 1, Device 2 and Device 3.}
	\label{fig4}
	\vspace{-0.1in}
\end{figure}
\begin{figure}[ht]
	\vspace{-0.1in}
	\centering
	\includegraphics[width=0.41\textwidth]{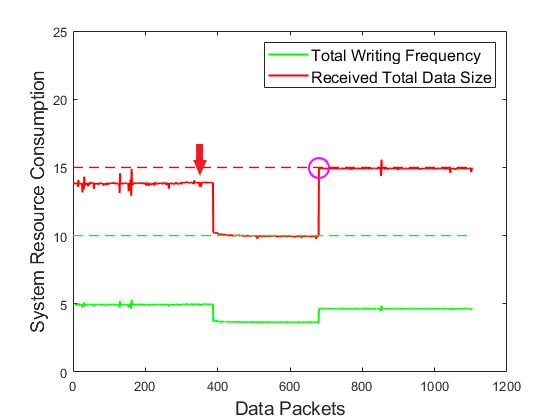}
	\caption{System resource consumption in boundary conditions. The magenta cycle shows the connection point of device 3.}
	\label{fig5}
	\vspace{-0.1in}
\end{figure}

Based on the simulation platform, the decentralized optimization process and system resource usage are shown in Fig. \ref{fig4} and Fig. \ref{fig5} in the scenario of insufficient resources. We note that before the connection of device $3$, device $1$ and device $2$ transmit their data packets under the corresponding optimized transmission frequencies exactly as described in the first scenario with sufficient resources. As shown in Fig. \ref{fig4}, after the device $3$ connects to the system (indicated by the red arrow), the DFWF of device $2$ is readjusted and converges to a new optimal value. The DFWF of device $1$ remains unchanged since the recalculated optimal result equals the previously assigned DFWF before the connection of device $3$. After the decentralized ADMM solution is found for device $3$ (indicated by the magenta circle), device $3$ pushes data packets to the gateway using its optimal DFWF. After all three devices are transmitting data steadily (i.e., after the magenta circle), our results show that the estimated DFWFs are $0.9984\ Hz$, $2.6410\ Hz$ and $0.9984\ Hz$ for device $1$, $2$, and $3$, respectively, which are reported in Table \ref{simulationB}. Again, these estimated DFWFs are slightly below the theoretical optimal DFWFs, indicated by dotted-lines, reflecting time delays of $1.6\ ms$, $3.7\ ms$ and $1.6\ ms$ for devices $1$, $2$, $3$, respectively during their transmissions.

After the optimal transmission frequencies are established, as shown in Fig. \ref{fig5}, device $3$ starts to push data (marked by the magenta circle) and the total writing data size reaches the level of the system resource boundary immediately. This indicates that the proposed system is able to reallocate the system resources to finish the data transmission task effectively using the ADMM approach. Finally, for comparison purposes, we evaluate the overall utility under the ADMM-optimized DFWFs, with non-optimized average distributed DFWFs (i.e., $x_i = c/N$), and non-optimized proportionally distributed DFWFs (i.e., $x_i = (a_i*c)/\sum a_i$) as two baselines given the same MWF $c$. The results shown in Table \ref{UtilityValue} find that the utility under ADMM-optimized DFWFs achieves the largest value, which demonstrates that the proposed system obtains the best result compared to other trivial system setups that have not undergone any optimization process. 

\begin{table}[!h]
	\centering
	\caption{Simulation results (average)}\label{simulationB}
	\vspace{-0.2cm}
	\begin{tabular}{|c|c|c|c|}
		\hline
		\textbf{ } & \textbf{DFWF (Hz)} & \textbf{DFWF (Hz)} & \textbf{DFWF (Hz)}  \\ \hline
		\textbf{ } & \textbf{Device 1} & \textbf{Device 2} & \textbf{Device 3}  \\ \hline
		\textbf{Theoretical} & 1.0000 & 2.6667 & 1.0000  \\ \hline
		\textbf{Actual} & 0.9984 & 2.6410 & 0.9984  \\ \hline
		\textbf{Absolute Error} & 0.0016 & 0.0257 & 0.0016 \\ \hline
	\end{tabular}
	\vspace{-0.2in}
\end{table}

\begin{table}[ht!]
	\centering
	\caption{Utility Evaluation}\label{UtilityValue}
	\vspace{-0.2cm}
	\begin{tabular}{|c|c|}
		\hline
		\textbf{DFWFs} &
		\textbf{Utility Value} \\ \hline
		ADMM optimized &  1381.22 \\ \hline
		Average distributed & 1190.35 \\ \hline
		Proportionably distributed & 1086.00 \\ \hline	
	\end{tabular}
	\vspace{-0.1in}
\end{table}

\begin{figure}[ht]
	\vspace{-0.1in}
	\centering
	\includegraphics[width=0.41\textwidth]{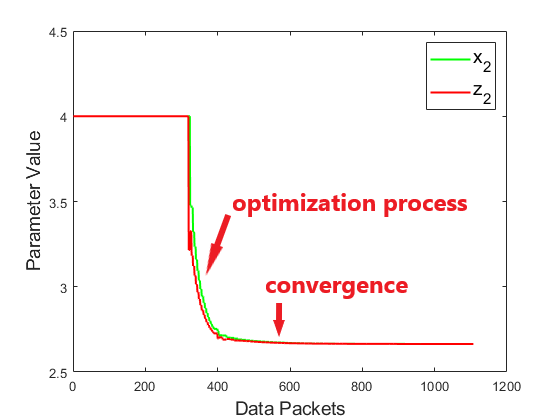}
	\caption{Monitoring of $z_{2}$ and $x_{2}$ during the decentralized optimization process for Device 2.}
	\label{fig6}
	\vspace{-0.2in}
\end{figure}

\section{Abnormal transmission frequency detection}

While the transmission frequencies are determined and allocated by the system, all the devices push data steadily with their specified DTWF. However, an IoT device may transmit data with an unexpected transmission frequency when an edge device intensively manipulates its converged DFWF to a different, more desired, state. Alternatively, a malicious node in the network may tamper with the converged DFWF of an edge device intentionally. In this section, manipulation of transmission frequencies is briefly discussed for the examination of abnormal transmission frequency detection at the side of gateway.

According to the fundamental mechanism of the ADMM algorithm, the gateway only has access to $\textbf{z}$. Since $\textbf{x}$ achieves convergence to $\textbf{z}$ eventually, as a specific example (i.e., $z_{2}$ and $x_{2}$) shown in Fig. \ref{fig6}, we argue that the gateway is able to detect the anomaly of $\textbf{x}$ during the whole transmission process based on its knowledge of the latest value of $\textbf{z}$. Specifically, this detection process can be described in the following three steps:

\begin{itemize}
	\item[S1:] Gateway accesses the value of $z_{i}$ for each device.
	\item[S2:] Gateway estimates the DFWF (i.e., the converged value of $x_{i}$) for each device according to the received time-stamped data flow.
	\item[S3:] If the estimated DFWF is significantly different to the reference value of $z_{i}$ (i.e., $|z_{i}-x_{i}| \geq \delta$, where $\delta$ is a threshold depending on the network delay), the optimal transmission frequency can regarded as anomalous and as being manipulated.
	
\end{itemize}

\section{Conclusion}\label{sec:Conclusion}

In this paper, we propose a novel transmission frequency management system for IoT edge devices. This innovative system is able to find the optimal transmission frequency for each IoT device in a resource-constrained, privacy-aware environment. In addition, it is able to detect the connection of a new device and determine and reassign the new optimal transmission frequencies automatically. Our simulation results show that the proposed system is effective in real-world scenarios, with a high accuracy for estimation of transmission frequency in a low-latency ($5\ ms$) router-based experimental IoT network. Finally, we have introduced an abnormal frequency detection mechanism for simple scenarios where the converged DFWF may have been manipulated. Our results show that the ADMM-based algorithm can successfully identify this type of undesirable anomaly during real-time IoT data transmission. 

As part of our future work, different kinds of abnormal frequency detection mechanisms will be further investigated by taking account of more complicated application scenarios where malicious manipulations of utility functions of devices are considered.  We will also investigate the dynamics of system behaviours when the utility functions are non-smooth and non-convex.


\section*{Acknowledgement}

This work was support in part by Science Foundation Ireland grant SFI/12/RC/2289\_P2, and in part by Postgraduate Research Scholarship from the Faculty of Engineering and Computing at Dublin City University. 

\bibliographystyle{ieeetran}
\bibliography{References}

\end{document}